\documentclass[aip,jcp,reprint]{revtex4-2}
\usepackage{amsmath, amssymb, physics}
\usepackage{graphicx}
\usepackage{bm}
\usepackage{tikz}
\usepackage[version=3]{mhchem} 
\usepackage{subfigure}

\usetikzlibrary{arrows.meta}    
\usetikzlibrary{decorations.pathmorphing} 
\usetikzlibrary{patterns}       
\usetikzlibrary{positioning}    
\usetikzlibrary{calc}           

\begin{document}

\title{Fractional Statistics and Electron Transfer at Topological Defects
}

\author{Eric~R.~Bittner}
\email{ebittner@central.uh.edu}
\affiliation{Department of Physics, University of Houston, Houston, Texas 77204, United~States}

%

\begin{abstract}
We develop a theoretical framework for electron transfer (ET) at graphene defects, treating the surface as a Dirac cone with a localized defect state coupled to a vibrational environment. 
Using a polaron transformation combined with a modified density of states, we derive an explicit expression for the ET rate that incorporates both vibrational reorganization and fractionalized quasiparticle statistics.
We show that fractional statistics, modeled through a power-law density of states, suppress low-energy ET near resonance and introduce tunable deviations from conventional Marcus-like kinetics.
Our results suggest that strain, defect engineering, or chemical modification could stabilize fractional excitations in graphene-based catalysts, offering new strategies for controlling surface reactivity.
These findings provide a foundation for future experimental and computational investigations into the role of topology and fractional statistics in chemical electron transfer.
\end{abstract}

\maketitle
\section{Introduction}

Controlling electron transfer (ET) processes with high specificity and resilience remains a central goal in physical chemistry and catalysis. Recent advances in materials science have introduced a new class of systems—topological insulators—that offer novel mechanisms for achieving robustness against disorder, decoherence, and environmental fluctuations.

Topological insulators are materials that are insulating in the bulk yet host conducting states at their boundaries. These boundary (edge) states are protected by global topological invariants of the system’s electronic structure and are remarkably resilient to local perturbations such as defects, chemical functionalization, or thermal disorder. This intrinsic stability suggests a new design principle for catalysts: \emph{topological catalysis}, in which topologically protected electronic states mediate charge transfer processes with enhanced robustness and efficiency.

In two-dimensional (2D) topological insulators, edge states can often be modeled as one-dimensional (1D) Dirac fermions, exhibiting a linear energy-momentum relation and spin-momentum locking. This simple but powerful description provides a natural framework for analyzing ET reactions mediated by topological edge states.
Recent theoretical studies have suggested an even more intriguing possibility: that defects in graphene can host excitations with fractional statistics, or \emph{anyons}.\cite{Seradjeh2008, Obispo2015, Ryu2009} 
Anyons interpolate between bosons and fermions, acquiring a statistical phase $e^{i\theta}$ under exchange, where $\theta$ differs from the conventional $0$ (bosons) or $\pi$ (fermions). Anyonic excitations, with $\theta = \pi/2$, have been proposed in certain fractional quantum Hall states and lattice models of topological order. \cite{Veillon2024,Glidic2024}
Unlike Majorana fermions, which are self-conjugate but obey ordinary fermionic statistics ($\theta = \pi$), anyons represent a truly fractionalized form of quantum statistics. Their unique exchange properties can influence dynamical processes such as tunneling, transport, and, as discussed here, surface ET rates through modifications of the local electronic density of states.  

Motivated by these developments, we develop an analytical model for ET into defect-functionalized graphene systems, incorporating both vibrational reorganization effects and potential fractionalization of the local electronic environment. Using a combination of polaron transformation and fractionalized density of states modeling, we derive explicit expressions for the ET rate, highlighting how topological protection, vibrational dressing, and fractional statistics intertwine to govern electron transfer dynamics.
\section{Minimal Model for Topological Catalysis}

We model the catalytic surface as a graphene sheet containing a single defect site located at the origin.
Under this assumption, a reactant molecule binds to the defect site, undergoes electron transfer with the graphene surface, and subsequently desorbs as reaction products.
The overall process can be represented schematically as
\begin{align}
\ce{A} + \ce{S} 
\xrightarrow{\text{ET}} 
\ce{[A-S]$^\ddagger$} 
\rightarrow 
\ce{A^* + S}
\end{align}
where \ce{A} is the reactant, \ce{S} denotes the surface defect site, 
\ce{[A-S]$^\ddagger$} represents the activated complex, and 
\ce{A$^*$} is the electronically excited or transformed product.
The electronic states near a graphene defect can be modeled as a Dirac cone centered at the K point of the Brillouin zone, 
since the low-energy excitations in pristine graphene arise from linear band crossings at these symmetry points, and local perturbations (e.g., vacancies or bond rearrangements) preserve this linear dispersion within a finite region of momentum space. This is a key assumption in our model.

\begin{figure*}[!t]
\subfigure[]{\includegraphics[width=0.4\textwidth]{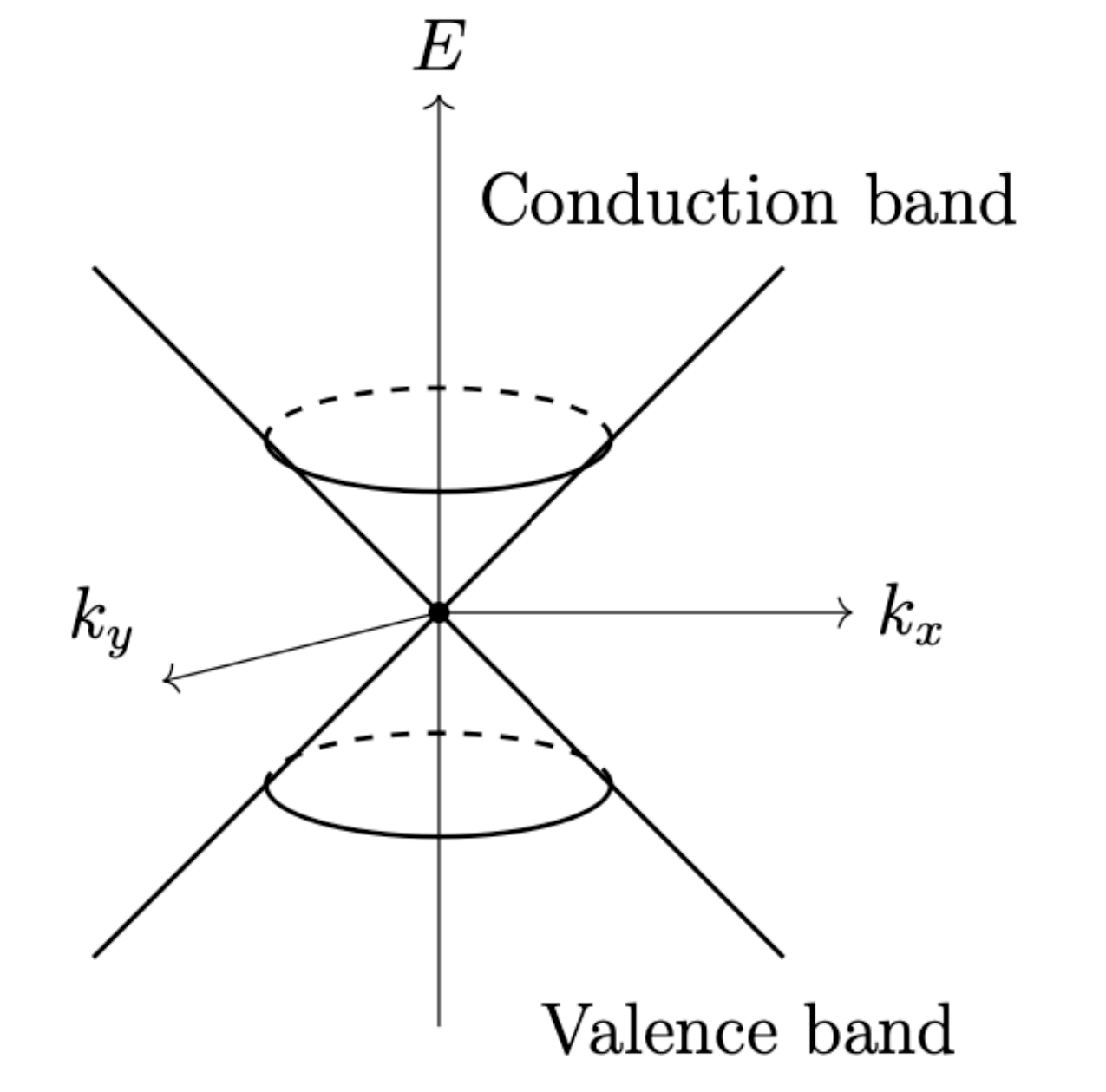}}
\subfigure[]{\includegraphics[width=0.4\textwidth]{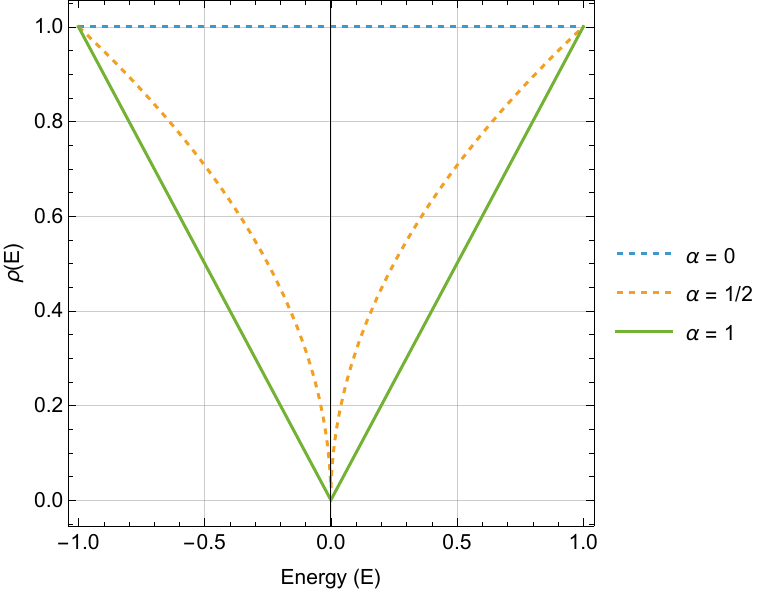}}
\caption{
(a) Schematic of the Dirac cone structure in graphene, illustrating the linear energy-momentum dispersion near the K point.
The $k_x$ and $k_y$ axes represent momentum components in the plane, and constant energy slices are shown by ellipses.
(b) Schematic plot of the electronic density of states $\rho(E)$ as a function of energy $E$ for different values of the fractional scaling exponent $\alpha$. The case $\alpha=1$ corresponds to the linear Dirac density of states, while smaller values of $\alpha$ reflect modifications associated with fractionalized quasiparticles near graphene defect sites.
}
\label{fig:dirac_cone}
\end{figure*}

Near the Dirac point, the energy varies linearly with the momentum $\mathbf{q}$ measured from the K point in the Brillouin zone,
\begin{equation}
E(\mathbf{q}) \approx \pm \hbar v_F |\mathbf{q}|,
\end{equation}
where $v_F$ is the Fermi velocity. The structure of the Dirac cone and its linear energy-momentum relation are illustrated schematically in Figure~\ref{fig:dirac_cone}(a).
This linear dispersion leads to a local density of states (LDOS) that vanishes at the Dirac point and increases linearly with energy,
\begin{equation}
\rho_{\text{Dirac}}(E) = \frac{|E|}{2\pi \hbar^2 v_F^2},
\end{equation}
with a slope determined by the Fermi velocity as shown in  Figure~\ref{fig:dirac_cone}(b).
This behavior contrasts sharply with that of a conventional two-dimensional electron gas (2DEG) with parabolic dispersion $E(\mathbf{q}) = \hbar^2 q^2 / 2m^*$, where $m^*$ is the effective mass.
For a 2DEG, the density of states is constant and energy-independent:
\begin{equation}
\rho_{\text{2DEG}} = \frac{m^*}{\pi \hbar^2}.
\end{equation}
Thus, graphene-like systems exhibit a qualitatively different energy dependence of the electronic density of states compared to traditional semiconductor materials.\cite{CastroNeto2009, Ando2005}
To connect to real materials, we can estimate $v_F$ for representative 2D systems.
For graphene, $v_F \approx 10^6~\text{m/s}$, consistent with experimental measurements.
Other 2D materials with topological edge states, such as stanene, bismuthene, or engineered heterostructures, are expected to exhibit comparable Fermi velocities within an order of magnitude.
In our model, we adopt $v_F = 10^6$~m/s as a representative value unless otherwise specified.

We next introduce a localized catalytic site at the defect, modeled as a discrete electronic orbital with energy $\epsilon_d$.
The Hamiltonian for the site is
\begin{equation}
H_{\text{site}} = \epsilon_d d^\dagger d,
\end{equation}
where $d^\dagger$ and $d$ are the creation and annihilation operators for the site orbital.
Electron transfer between the Dirac cone states and the catalytic site is mediated by a tunneling coupling:
\begin{equation}
H_{\text{coupling}} = \sum_{k}V_{k} \left( d^\dagger c_{k} + c^\dagger_{k} d \right) \approx  V (d^\dagger c_{0} + c^\dagger_{0} d)
\end{equation}
where $c_{k}$ annihilates an electron from the Dirac cone.  We assume local coupling to the continuum of Dirac states, with momentum-dependence 
absorbed into the energy-dependent density of states.
In the absence of vibrational coupling, the electron transfer rate from the Dirac cone to the catalytic site follows directly from Fermi's golden rule:
\begin{equation}
k_{\text{ET}} = \frac{2\pi}{\hbar} |V|^2 \rho(\Delta E),
\end{equation}
highlighting the role of the linear Dirac density of states in mediating the electron transfer process.
It is essential to clarify the meaning of the diabatic energy gap $\Delta E$ in the present model.
Although pristine graphene exhibits a vanishing band gap at the Dirac point, $\Delta E$ here refers not to an intrinsic electronic gap but to the energy difference between the reactant and product diabatic potential energy surfaces along the nuclear reaction coordinate as per Marcus theory. 
$\Delta E$ incorporates both the intrinsic reaction free energy $\Delta G^\circ$ and the reorganization energy $\lambda$ associated with structural rearrangements at the defect site.
As a result, despite the availability of low-energy electronic states in graphene, the overall electron transfer process remains activated, driven by nuclear reorganization.
We propose that chemical modification of defect structures or applying external mechanical strain can modify $\Delta E$, providing a potential lever for tuning ET kinetics.

\subsection{Thermal Effects and Full Quantum Rate Expression}

To incorporate vibrational coupling into the electron transfer (ET) process, we model the defect site as coupled to a local vibrational mode of frequency $\Omega$.
The polaron transformation provides a convenient framework for accounting for the influence of vibrational degrees of freedom on electronic transitions.
\cite{Pereverzev:2006aa}
We begin with the standard electron-phonon interaction Hamiltonian in the displaced oscillator model:
\begin{equation}
H_{\text{el-ph}} = \lambda\, d^\dagger d (a^\dagger + a),
\end{equation}
where \(\lambda\) is the coupling strength, and \(a^\dagger\), \(a\) are phonon creation and annihilation operators.
To account for vibrational reorganization, we apply a polaron (Lang-Firsov) transformation\cite{Wagner1986,LangFirsov1962}
\begin{equation}
U = \exp[g(a^\dagger - a)d^\dagger d], 
\end{equation}
with \(g = \lambda / (\hbar \Omega)\).
This leads to a transformed tunneling operator
\begin{equation}
\tilde{V} = U H_{\text{el-ph}} U^{\dagger} = V \exp\left[ g(a^\dagger - a) \right],
\end{equation}
where $a^\dagger$ and $a$ are the phonon creation and annihilation operators, and $g$ is the Huang-Rhys factor characterizing the coupling strength\cite{Pereverzev:2006aa}.

\begin{widetext}
The ET rate can be expressed in terms of the autocorrelation function of the dressed coupling operator:
\begin{equation}
k_{\text{ET}} = \frac{2}{\hbar^2} \mathrm{Re} \int_0^\infty dt\, e^{i \Delta E t/\hbar} \langle \tilde{V}(t) \tilde{V}(0) \rangle,
\end{equation}
where $\Delta E$ is the energy gap between the initial and final diabatic surfaces.
The autocorrelation function $C(t) = \langle \tilde{V}(t) \tilde{V}(0) \rangle$ can be evaluated exactly for a harmonic oscillator bath.
Using the time evolution of the phonon operators, we obtain\cite{Pereverzev:2006aa}
\begin{equation}
C(t) = V^2 \exp\left\{ -2g \left[ (1 - \cos \Omega t) \coth\left( \frac{\hbar \Omega}{2k_B T} \right) - i \sin \Omega t \right] \right\}.
\end{equation}

This expression retains full quantum and thermal dependence through the $\coth\left( \hbar \Omega/2k_B T \right)$ factor.
In the high-temperature limit, $k_B T \gg \hbar \Omega$, the hyperbolic cotangent simplifies to
\begin{equation}
\coth\left( \frac{\hbar \Omega}{2k_B T} \right) \approx \frac{2k_B T}{\hbar \Omega},
\end{equation}
yielding the approximate correlation function
\begin{equation}
C(t) \approx V^2 \exp\left\{ -\frac{4g k_B T}{\hbar \Omega} (1 - \cos \Omega t) + 2i g \sin \Omega t \right\}.
\end{equation}
The full rate expression can be evaluated by Fourier transforming $C(t)$.
Using the Jacobi-Anger expansion and performing the integration, the ET rate takes the form of a Franck-Condon weighted sum:
\begin{equation}
k_{\text{ET}} = \frac{2\pi}{\hbar} V^2 e^{-S \coth\left( \hbar \Omega / 2k_B T \right)} \sum_n I_n\left( \frac{S}{\sinh\left( \hbar \Omega / 2k_B T \right)} \right) \delta(\Delta E - n \hbar \Omega),
\end{equation}
where $S = 2g$ is the reorganization strength and $I_n$ is the modified Bessel function of the first kind.
This structure reflects the multiphonon nature of the electron transfer process: each term in the sum corresponds to a transition involving $n$ phonons, with thermally broadened Franck-Condon factors controlling the relative weights.
Importantly, even at high temperatures, thermal effects remain visible through the Gaussian envelope of the vibrational sidebands, ensuring that the rate expression maintains proper physical temperature dependence.
\end{widetext}

\subsection{Fractional Statistics and the Electron Transfer Rate}

Up until now, we have assumed that the electronic states near the graphene defect obey conventional Fermi-Dirac statistics, recent theoretical proposals suggest that localized defects in two-dimensional materials can host excitations with fractionalized exchange statistics.\cite{Seradjeh2008, Obispo2015, Ryu2009}
In the context of electron transfer, fractional statistics modify the local density of final states available for tunneling.
Rather than entering through the vibrational dressing, which remains bosonic, the effect of fractionalization is captured by a modification of the local electronic density of states (LDOS).
Several theoretical models, including chiral Luttinger liquids and edge states in fractional quantum Hall systems, yield power-law exponents of the form
\begin{equation}
\alpha(\theta) \sim \sin^2\left( \frac{\theta}{2} \right),
\end{equation}
ensuring that $\alpha = 0$ for bosons, $\alpha = 1$ for fermions, and fractional values in between.\cite{Wen1990, KaneFisher1992}
In particular, the LDOS is assumed to follow a power-law scaling of the form
\begin{equation}
\rho(E) \propto |E|^{\alpha} \Theta(E),
\end{equation}
where $\alpha$ is a statistics-dependent exponent.
(Fig.~\ref{fig:dirac_cone}(b)).

The presence of the step function $\Theta(\Delta E - n\hbar\Omega)$ ensures that only energy-conserving or physically allowed transitions contribute to the rate, thus preventing any divergence in the sum over vibrational sidebands. The emergence of fractional statistics at a graphene defect site could lead to measurable deviations from conventional Marcus-like electron transfer kinetics. In particular, the temperature and energy dependence of $k_{\text{ET}}$ would reflect the nontrivial scaling of the local density of states, offering a potential experimental signature of fractionalization. Future extensions of this model could incorporate more detailed statistical field theories to rigorously derive the dependence of $\alpha$ on defect structure and external strain.

\begin{figure}[ht]
    \centering
     \includegraphics[width=0.75\columnwidth]{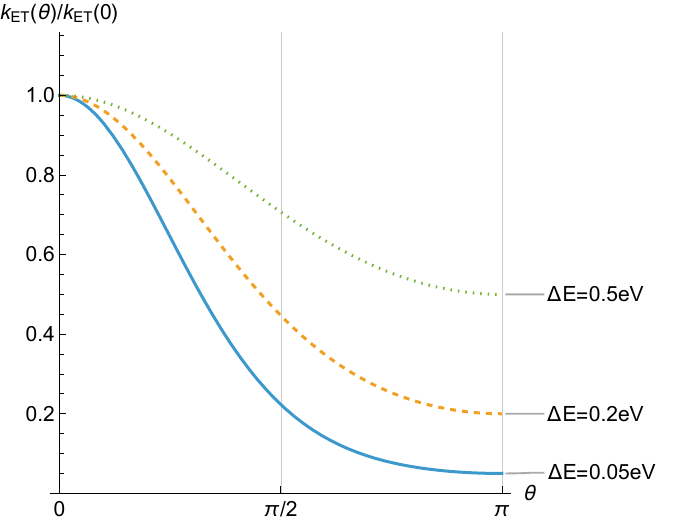}
    \caption{Dependence of the normalized electron transfer rate $k_{\text{ET}}(\theta)$ on the quasiparticle statistical angle $\theta$ for 
    different values of the energy gap $\Delta E$. The rate is normalized to its value at $\theta = 0$ for each $\Delta E$. Smaller energy gaps exhibit stronger suppression of the electron transfer rate with increasing $\theta$, reflecting the enhanced sensitivity to fractional statistics near resonance. For larger gaps, the effect of fractionalization becomes weaker, and the rate dependence flattens. The statistical exponent $\alpha$ varies as $\alpha(\theta) = \sin^2(\theta/2)$.
}
    \label{fig:rate_theta}
\end{figure}

Accordingly, the vibrationally dressed electron transfer rate is modified to
\begin{equation}
k_{\text{ET}} \propto \sum_n e^{-S} I_n(2g)\, |\Delta E - n\hbar \Omega|^{\alpha} \Theta(\Delta E - n\hbar \Omega),
\end{equation}
where $S = 2g$ is the polaron reorganization strength, and the Heaviside function enforces energy conservation and suppresses unphysical divergences.
This rate expression captures the essential influence of fractional statistics: as $\theta$ increases from $0$ to $\pi$, the corresponding increase in $\alpha$ suppresses low-energy electron transfer transitions.
This effect is most significant when the energy gap $\Delta E$ is small, i.e., when electron transfer occurs near resonance with a low-order vibrational sideband ($n \sim 0$ or $1$).
In this regime, the power-law scaling $|\Delta E - n\hbar \Omega|^\alpha$ introduces substantial suppression of the rate for $\alpha > 0$.
For example, if only the $n=0$ term contributes significantly, the rate behaves as
\begin{equation}
k_{\text{ET}}(\theta) \propto |\Delta E|^{\alpha(\theta)},
\end{equation}
which varies strongly with $\theta$ when $\Delta E$ is small.
Figure~\ref{fig:rate_theta} illustrates the normalized ET rate $k_{\text{ET}}(\theta)/k_{\text{ET}}(0)$ as a function of $\theta$ for several representative values of $\Delta E$.
As shown, the suppression due to fractional statistics is most pronounced near resonance and diminishes for larger energy gaps.

This behavior highlights a practical consequence of fractionalization:
electron transfer mediated by fractionalized quasiparticles becomes increasingly inefficient in the low-energy limit.
The dependence on $\theta$ could serve as an experimental signature of fractionalization if measurable deviations from standard Fermi golden rule kinetics are observed.
For instance, temperature-programmed measurements or bias-dependent conductance studies might reveal power-law trends consistent with a statistics-induced suppression of electronic transitions.


It is important to note that the specific form of the local electronic density of states $\rho(E)$ near a graphene defect may vary depending on the nature of the defect, the degree of strain, and the presence of external fields or interactions. While the power-law scaling $\rho(E) \propto |E|^{\alpha}$ provides a generic framework for capturing the influence of fractionalized excitations, the precise value of the exponent $\alpha$ could differ across different types of defects. In particular, defects inducing localized mid-gap states, pseudo-magnetic fields, or topological textures may exhibit additional features such as subgap resonances or asymmetries in $\rho(E)$. A more detailed microscopic analysis of the defect-induced electronic structure, potentially via density functional theory (DFT) or tight-binding calculations incorporating fractionalization effects, would provide a more accurate basis for modeling electron transfer rates in specific systems. Exploring such defect-specific variations remains an exciting direction for future theoretical and experimental investigation.

\section{Summary and Outlook}

We have developed a minimal theoretical framework for electron transfer (ET) from topologically protected electronic states at graphene defects, incorporating both vibrational reorganization and the possibility of fractionalized quasiparticles.
By combining a polaron transformation with a statistics-dependent local density of states, we derived explicit expressions for the ET rate, revealing how fractional statistics suppress electron transfer near resonance.

Our results suggest that strain, defect engineering, or chemical modification of graphene could stabilize electronic excitations with nontrivial exchange statistics, leading to tunable and possibly robust catalytic behavior.
The interplay between vibrational dressing and fractional statistics introduces a novel mechanism for controlling ET kinetics that goes beyond conventional Marcus-like or Fermi golden rule descriptions.
While previous studies have explored the impact of fractional nuclear dynamics on electron transfer kinetics, particularly in biological and polymeric systems,\cite{Goychuk2019} our work focuses on a distinct mechanism.
Here, the modification arises not from subdiffusive nuclear motion, but from the fractionalization of the electronic density of states at topological defects.
This distinction offers a complementary perspective on how nontrivial quantum statistics can influence surface electron transfer reactions.

Future theoretical work could extend this framework to include many-body effects, dynamic disorder, or explicit simulation of surface reactions under strain or external fields.
Experimentally, probing the energy or temperature dependence of ET rates, or selectively engineering defect environments, could provide signatures of fractionalization in realistic catalytic systems.
Synthetic approaches to designing graphene-based or related 2D catalysts with targeted topological or strain profiles offer exciting possibilities for harnessing fractional statistics in chemical reactivity.

\section*{Acknowledgments}
This work was supported by the National Science Foundation (CHE-2404788) and the Robert A. Welch Foundation (E-1337).

\bibliography{references}      

\end{document}